\documentclass[brazil,english,aps,showpacs]{revtex4}
\usepackage[T1]{fontenc}
\usepackage[latin9]{inputenc}
\usepackage{amsmath}
\usepackage{amssymb}
\usepackage{esint}

\makeatletter
\@ifundefined{textcolor}{}
{%
 \definecolor{BLACK}{gray}{0}
 \definecolor{WHITE}{gray}{1}
 \definecolor{RED}{rgb}{1,0,0}
 \definecolor{GREEN}{rgb}{0,1,0}
 \definecolor{BLUE}{rgb}{0,0,1}
 \definecolor{CYAN}{cmyk}{1,0,0,0}
 \definecolor{MAGENTA}{cmyk}{0,1,0,0}
 \definecolor{YELLOW}{cmyk}{0,0,1,0}
 }

\makeatother

\usepackage{babel}

\begin{document}

\title{Considerations on the Graviton Excitation Modes of Ho\v{r}ava-Lifshitz
Gravity}

\author{B. Pereira-Dias}

\email{bpdias@cbpf.br}

\affiliation{Centro Brasileiro de Pesquisas Físicas (CBPF),\\
Rua Dr. Xavier Sigaud 150, Urca,\\
 Rio de Janeiro, RJ, Brazil, CEP 22.290-180}

\author{C. A. Hernaski}

\email{carlos@cbpf.br}

\affiliation{Universidade Federal do Amapá (Unifap),\\
Rod. Juscelino Kubitschek KM-02, Jardim Marco Zero,\\
 Macapá, AP, Brazil, CEP 68.902-280}

\author{J. A. Helayël-Neto}

\email{helayel@cbpf.br}

\affiliation{Centro Brasileiro de Pesquisas Físicas (CBPF),\\
Rua Dr. Xavier Sigaud 150, Urca,\\
 Rio de Janeiro, RJ, Brazil, CEP 22.290-180}
\begin{abstract}
A new set of projection operators is constructed to suitably handle
non-relativistic theories of gravity with anisotropic scaling, including
the ones with parity-violating terms. This alternative procedure allows
us to discuss unitarity and spectral properties for different formulations
of the Ho\v{r}ava-Lifshitz gravity. This task, that generally involves
lengthy algebraic steps, becomes more systematic and greatly simplified
in terms of the projectors we work out. Moreover, this procedure allows
us to fix the number of propagating degrees of freedom and the structure
of gauge symmetries is readily determined. In order to test the efficacy
of the technique at hand, the unitarity and low-energy regime of a
general Ho\v{r}ava-Lifshitz Gravity model are investigated. 
\end{abstract}



\maketitle

\section{Introduction}

The recent proposal of Petr Ho\v{r}ava for a new theory of gravity
\cite{Horava:2008ih,Horava:2009uw,Horava:2010zj,Horava:2011gd} was
received with great enthusiasm by the Field Theory, Gravitation and
Cosmology communities \cite{Bemfica:2011a,Bemfica:2011b,Blas:2009qj,Blas:2009yd,Blas:2010hb,Blas:2011zd,Bogdanos:2009,Borzou:2011mz,Cai:2010hi,Calcagni:2009qw,Charmousis:2009tc,Garfinkle:2011iw,Gumrukcuoglu:2011ef,Gumrukcuoglu:2011xg,Herdeiro:2011,Jacobson:2010mx,Koyama:2009hc,Li:2009bg,Lin:2011,Myung:2010uq,Nastase:2009nk,Ong:2011km,Padilla:2010ge,Papazoglou:2009fj,Pospelov,daSilva:2010bm,Sotiriou:2009bx,Sotiriou:2009gy,Sotiriou:2010wn,Takahashi:2009wc,Visser:2009fg,Visser:2011mf,Weinfurtner:2010hz,Wang:2010,Zhu:2011}
and interesting consequences for cosmological models have been worked
out \cite{Brandenberger:2009yt,Dutta:2010,Furtado:2011tc,Gao:2009ht,Saridakis:2011,Mokohyama:2010xz,Wang:2009,Wang:2009b}.
The large appraisal for this model, dubbed as Ho\v{r}ava-Lifshitz
gravity (HLG), came from the fact that it is a candidate for a consistent
quantum field theory for gravity, since it has a remarkably improved
ultraviolet behavior. However, to achieve such a result, Ho\v{r}ava
had to give up diffeomorphism invariance as a fundamental symmetry
so to impose \emph{anisotropic scaling} of the space and time dimensions.

Despite its success, it soon became evident that Ho\v{r}ava's original
formulation including the \emph{detailed balance condition }is not
obviously a phenomenologically viable model \cite{Charmousis:2009tc,Sotiriou:2009bx,Sotiriou:2009gy}.
So, to reconcile HLG with experimentation, it became necessary to
loose up such condition and allow the inclusion of a large number
of Lagrangian terms with arbitrary coefficients in the model or introduce
auxiliary fields to enlarge the symmetries of the model. Besides this,
HLG can be formulated in two distinct versions: \emph{nonprojectable}
or \emph{projectable} -- which refers to the dependence or not of
the lapse function with the space coordinates. These different possible
formulations brought a discussion in the literature concerning whether
each of these formulations should be considered as a \emph{good},
\emph{bad}, \emph{healthy}, or \emph{ugly} version of HLG \cite{Blas:2010hb,Padilla:2010ge}.

All this debate included not only analysis of the phenomenological
aspect, but also considerations on the the quantum consistency of
each version of theory. It is in this context that we judge necessary
(and convenient) to develop a general method for the attainment of
the propagators and, thereby, the identification of the excitations
present in the possible models based on the HLG frame. With the propagators
at hand it is possible to systematically obtain a description of the
particle spectrum and the relativistic and quantum properties of scattering
processes. More exactly, one is able to pinch out the propagating
modes and determine conditions on the coefficients of the Lagrangian
so as to impose the absence of tachyons and ghosts.

There are several methods for the attainment of propagators, but,
particularly in the case of weak field approximation for quantum gravity,
which is our interest, algebraic methods have been intensively developed,
specially the one based on the spin projection operators (SPO) \cite{rivers1964,neville,S-N}.
The SPO for Lorentz invariant models has the interesting property
of being a orthonormal basis that decomposes the fields into definite
spin-parity sectors. In this paper we follow the lines of previous
works \cite{Hernaski:2009wp,extending the spin projection operators,Chern-Simons gravity,PereiraDias:2010gb}
to build up an orthonormal basis of operators that is specially suitable
to handle of HLG models, including the ones with parity-breaking terms.
With this result we may discuss the spectral properties of the HLG
and understand better the extra 3D spin-0 propagating mode that appear
in the theory. Moreover, this technology allows us to determine the
number of propagating degrees of freedom and the gauge symmetry structure
is easily determined. We would like to point out that recently F.
S. Bemfica and M. Gomes also discuss the propagators and consistency
conditions on the spectrum for special classes of HLG models by adopting
an independent method \cite{Bemfica:2011a,Bemfica:2011b}.

This paper is organized as follows: In Sec. \ref{sec:Decomposing-the-Graviton},
we build up the basis of projection operators suitable with the symmetries
of HLG. We present in the Sec. \ref{sec:Consistency-Analysis-of}
a general method for the consistency analysis of the particle spectrum.
The Lagrangian model of our interest shall be described in the Sec.
\ref{sec:Description-of-the} and the attainment of its propagators
along with the spectral analysis is done is Sec. \ref{sec:Attainment-of-The}.
We present our Concluding Assessment in Sec. \ref{sec:Concluding-remarks}
emphasizing in some consequences of the propagator structure of HLG
in the low energy regime. In the Appendix \ref{sec:Inverse-of-A0},
we list the inverse of the spin-0 matrix used to calculate the propagators
of Sec. \ref{sec:Attainment-of-The}, whereas the Appendix \ref{sec:Degree-of-Freedom-Projection-Operators}
is aimed to derive some useful algebraic relations that the projection
operators satisfy.

Unless stated otherwise, we shall use Latin letters ($a,b,c,\ldots$)
for 3D spatial index.  We also define $\delta_{ab}=\mbox{diag}\left(1,1,1\right)$
and $\epsilon_{abc}$ as the completely antisymmetric symbol normalized
to $\epsilon_{012}=+1$. Also, we shall follow the conventions: $R_{\ bcd}^{a}=\partial_{c}\Gamma_{bd}^{a}+\Gamma_{ce}^{a}\Gamma_{bd}^{e}-\left(c\leftrightarrow d\right)$,
$\Gamma_{bc}^{a}=\frac{1}{2}g^{ad}\left(\partial_{b}g_{dc}+\partial_{c}g_{db}-\partial_{d}g_{bc}\right)$,
$R_{bd}=R_{\ bad}^{a}$, $R=g^{bd}R_{bd}$, with $g_{ab}$ being the
spatial metric.

\section{Decomposing the Graviton Excitation Modes\label{sec:Decomposing-the-Graviton}}

In the Ho\v{r}ava-Lifshitz gravity (HLG), the breaking of $\left(1+3\right)$-dimensional
diffeomorphisms invariance is implemented by endowing space-time with
a preferred foliation of three-dimensional spacelike surfaces. This
defines the splitting of coordinates into space and time that explicitly
breaks general covariance down to the subgroup of coordinate transformations\begin{equation}
t\mapsto\tilde{t}\left(t\right),\quad x^{a}\mapsto\tilde{x}^{a}\left(t,x^{a}\right),\label{eq:fDiffs}\end{equation}
referred in the literature as \emph{foliation-preserving diffeomorphisms}.

To fulfill the task of defining a Lagrangian model  with such a symmetry,
it is natural to make use of the ADM (Arnowitt-Deser-Misner) decomposition,
which in the convention we are adopting, decomposes the space-time
metric $^{\left(4\right)}g_{\mu\nu}$ in terms of the lapse $\mathcal{N}$,
the shift $N_{a}$, and the spatial metric $g_{ab}$ as \begin{equation}
ds^{2}=\left(\mathcal{N}^{2}-N_{a}N^{a}\right)dt^{2}-2N_{a}dx^{a}dt-g_{ab}dx^{a}dx^{b}.\label{eq:ADM-metric}\end{equation}
From the equation \eqref{eq:ADM-metric} we may relate the inverse
of the four-dimensional metric and the ADM fields by \begin{equation}
^{\left(4\right)}g^{00}=\frac{1}{\mathcal{N}^{2}},\quad{}^{\left(4\right)}g^{0a}=-\frac{N^{a}}{\mathcal{N}^{2}},\quad{}^{\left(4\right)}g^{ab}=-g^{ab}+\frac{N^{a}N^{b}}{\mathcal{N}^{2}},\end{equation}
and $\sqrt{-\det\negthinspace^{\left(4\right)}g_{\mu\nu}}=\mathcal{N}\sqrt{\det g_{ab}}$.

Having defined the fundamental fields of the model, we may perform
a weak field expansion around a Minkowski metric solution in order
to study the graviton excitation modes of HLG,\begin{equation}
\mathcal{N}=1+n,\quad g_{ab}=\delta_{ab}+h_{ab},\end{equation}
with $n$ and $h_{ab}$ being the quantum fluctuation of the lapse
function and the spatial metric, respectively, and the shift function
$N_{a}$ being the quantum fluctuation itself.

Breaking down the Minkowski symmetry, $SO\left(1,3\right)$, of the
linearized theory to a $SO\left(3\right)$ has some nontrivial consequences,
specially in the interpretation of the propagating modes. An usual
relativistic quantum field theory is invariant under Poincaré transformations.
It provides a classification for particles as unitary irreducible
representations of the Poincaré group by the quantum numbers: mass
and spin. In particular, the spin is characterized by the unitary
representations of the $SO\left(1,3\right)$ little group, that is
the subgroup of the Lorentz group that leaves the representative four-momentum
unchanged.

For the class of HLG models, we may proceed in a strict analogy. We
split our propagating modes in {}``what would be the spin'' defined
the subgroup of the rotation group which leaves the representative
three-momentum invariant. So, to arrange the degree-of-freedom into
these {}``spin modes'', we may develop a complete set of projection
operators for Ho\v{r}ava-like models. For this task, it is reasonable
to make use of the already developed three-dimensional $SO\left(1,2\right)$
Barnes-Rivers projection operators \cite{Hernaski:2009wp} and adapt
them for a $SO\left(3\right)$-invariant model implementing the following
substitution:\begin{equation}
\eta_{\hat{\mu}\hat{\nu}}\rightarrow\delta_{ab},\quad k_{\hat{\mu}}\rightarrow k_{a}\end{equation}
where $\hat{\mu},\hat{\nu}=0,1,2$, $a,b=1,2,3$, $k_{\hat{\mu}}$
is the (1+2)-D relativistic momentum and $k_{a}$ is the 3D spatial
momentum. For example, in the given situation, the transverse ($\theta_{ab}$)
and the longitudinal ($\omega_{ab}$) operators, which are building
blocks operators become\begin{subequations}\begin{eqnarray}
\theta_{ab} & = & \delta_{ab}-\omega_{ab},\\
\omega_{ab} & = & \frac{k_{a}k_{b}}{k^{2}},\end{eqnarray}
\end{subequations} with $k^{2}=k_{a}k_{a}$.

In this way, we may organize the HLG projection operators in a matrix-form
according to the respective spin sector ($P\left(0\right)$, $P\left(1\right)$,
and $P\left(2\right)$) to cover the case of parity-preserving Lagrangians
containing the fields $n$, $N_{a}$, and $h_{ab}$. They are arranged
by the possible spin representations carried in the quantum fields
-- $n\in\underbar{0}$, $N_{a}\in\underbar{0}\oplus\underbar{1}$,
$h_{ab}\in\underbar{0}\oplus\underbar{1}\oplus\underbar{2}$, which
comes from usual group-theoretic argumentation. So, we cast them as

\begin{subequations}\begin{align}
P(0) & =\negthickspace\begin{array}{c}
h_{ab}\\
h_{ab}\\
N_{a}\\
n\end{array}\negthickspace\overset{h_{cd}\qquad\ \ \ h_{cd}\qquad\ \ \ \ N_{c}\qquad\ \ \ n}{\left[\begin{array}{cccc}
\frac{1}{2}\theta_{ab}\theta_{cd} & \frac{1}{\sqrt{2}}\theta_{ab}\omega_{cd} & \frac{1}{\sqrt{2}}\theta_{ab}\bar{k}_{c} & \frac{1}{\sqrt{2}}\theta_{ab}\\
\frac{1}{\sqrt{2}}\omega_{ab}\theta_{cd} & \omega_{ab}\omega_{cd} & \omega_{ab}\bar{k}_{c} & \omega_{ab}\\
\frac{1}{\sqrt{2}}\bar{k}_{a}\theta_{bc} & \bar{k}_{a}\omega_{bc} & \omega_{ab} & \bar{k}_{a}\\
\frac{1}{\sqrt{2}}\theta_{cd} & \omega_{cd} & \bar{k}_{b} & 1\end{array}\right]}\label{eq:P0-spin}\end{align}

\begin{equation}
P\left(1\right)=\negthickspace\begin{array}{c}
h_{ab}\\
N_{a}\end{array}\negthickspace\overset{h_{cd}\qquad\qquad\qquad\qquad\qquad\qquad N_{c}}{\left[\begin{array}{cc}
\frac{1}{2}\left(\theta_{ac}\omega_{bd}+\theta_{bc}\omega_{ad}+\theta_{ad}\omega_{bc}+\theta_{bd}\omega_{ac}\right) & \negthickspace\frac{1}{\sqrt{2}}\left(\theta_{ac}\bar{k}_{b}+\theta_{bc}\bar{k}_{a}\right)\negthickspace\\
\frac{1}{\sqrt{2}}\left(\theta_{ba}\bar{k}_{c}+\theta_{ca}\bar{k}_{b}\right)\negthickspace & \theta_{ab}\end{array}\right]}\label{eq:P1-spin}\end{equation}

\begin{equation}
P^{hh}\left(2\right)=\frac{1}{2}\left(\theta_{ac}\theta_{bd}+\theta_{ad}\theta_{bc}-\theta_{ab}\theta_{cd}\right)\label{eq:P2-spin}\end{equation}
\end{subequations}where we have defined $\bar{k}_{a}=\frac{k_{a}}{\sqrt{k^{2}}}.$

However, even in Ho\v{r}ava's original proposal with \emph{detailed
balance} and its subsequent modifications there appear parity violating
terms, that notably cannot be written in terms of the $SO\left(3\right)$
Barnes-Rivers operators. A hint to circumvent this difficulty is to
note that in the presence of parity violating terms, the transverse
modes may propagate independently and not as parity-doublets (as discussed
in Ref. \cite{extending the spin projection operators} and put into
test in Ref. \cite{Chern-Simons gravity}). In practice, the two degrees
of freedom of the $\theta$-operator can be split by two orthonormal
operators,\begin{equation}
\theta_{ab}=\rho_{ab}+\sigma_{ab}\end{equation}
satisfying $\rho^{2}=\rho$, $\sigma^{2}=\sigma$, and $\rho\cdot\sigma=0$.
This decomposition is useful to handle with parity violating term
such as the ones containing the $\epsilon$-symbol, since $\rho$
and $\sigma$ may be defined to satisfy \begin{equation}
\epsilon^{abc}k_{c}=i\sqrt{k^{2}}\left(\rho_{ab}-\sigma_{ab}\right).\end{equation}

The degree-of-freedom projection operators for relativistic planar
models ($SO\left(1,2\right)$-invariant theories) with parity-breaking
terms have been thoroughly worked out in the Ref. \cite{extending the spin projection operators}
and can be used adequately for HLG models with parity violating terms.
Specializing these projection operators for the fields of the ADM
decomposition, one obtains,

\begin{subequations}\begin{equation}
P(1)=\negthickspace\begin{array}{c}
h_{ab}\left(+\right)\\
h_{ab}\left(-\right)\\
N_{a}\left(+\right)\\
N_{a}\left(-\right)\end{array}\negthickspace\overset{h_{cd}\left(+\right)\qquad\qquad h_{cd}\left(-\right)\qquad\qquad\qquad N_{c}\left(+\right)\qquad\qquad N_{c}\left(-\right)}{\left[\negthickspace\begin{array}{cccc}
2\rho_{ac}\omega_{bd}\negthickspace & \negthickspace2\epsilon_{ghe}\rho_{a}^{g}\sigma_{c}^{h}\omega_{bd}\bar{k}^{e}\negthickspace & \negthickspace\sqrt{2}\rho_{ac}\bar{k}_{b}\negthickspace & \negthickspace\sqrt{2}\epsilon_{ghe}\rho_{ag}\sigma_{c}^{h}\omega_{be}\\
2\epsilon_{ghe}\sigma_{a}^{h}\rho_{c}^{g}\omega_{db}\bar{k}^{e}\negthickspace & \negthickspace2\sigma_{ac}\omega_{bd}\negthickspace & \negthickspace\sqrt{2}\epsilon_{ghe}\sigma_{a}^{h}\rho_{c}^{g}\omega_{eb}\negthickspace & \negthickspace\sqrt{2}\sigma_{ac}\bar{k}_{b}\\
\sqrt{2}\rho_{ba}\bar{k}_{c}\negthickspace & \negthickspace\sqrt{2}\epsilon_{ghe}\sigma_{b}^{h}\rho_{a}^{g}\omega_{ec}\negthickspace & \negthickspace\rho_{ab}\negthickspace & \negthickspace\epsilon^{fgh}\rho_{af}\sigma_{bg}\bar{k}_{h}\\
\sqrt{2}\epsilon_{ghe}\rho_{bg}\sigma_{a}^{h}\omega_{ce}\negthickspace & \negthickspace\sqrt{2}\sigma_{ba}\bar{k}_{c}\negthickspace & \negthickspace-\epsilon^{def}\sigma_{ad}\rho_{be}\bar{k}_{f}\negthickspace & \negthickspace\sigma_{ab}\end{array}\negthickspace\right]}\label{eq:P1-dof}\end{equation}
\begin{equation}
P(2)=\negthickspace\begin{array}{c}
h_{ab}\left(+\right)\\
h_{ab}\left(-\right)\end{array}\negthickspace\overset{\qquad h_{cd}\left(+\right)\qquad\qquad\qquad\qquad\qquad\qquad\qquad h_{cd}\left(-\right)}{\left[\begin{array}{cc}
\frac{1}{2}(\rho_{ad}\rho_{bc}+\sigma_{ad}\sigma_{bc}-\rho_{ab}\sigma_{cd}-\sigma_{ab}\rho_{cd}) & \epsilon_{ghe}(\rho_{ca}\sigma_{d}^{h}\rho_{b}^{g}-\sigma_{cb}\rho_{d}^{g}\sigma_{a}^{h})\bar{k}^{e}\\
\epsilon_{ghe}(\rho_{ac}\sigma_{b}^{h}\rho_{d}^{g}-\sigma_{ad}\rho_{b}^{g}\sigma_{c}^{h})\bar{k}^{e} & 2\rho_{ac}\sigma_{bd}\end{array}\right]}\label{eq:P2-dof}\end{equation}
\end{subequations}In the above matrices it is understood the operators
share the same symmetrization properties (with numerical factor) of
the associated fields. 

It is noteworthy to emphasize that all the HLG projection operators
\eqref{eq:P0-spin}-\eqref{eq:P2-spin} and \eqref{eq:P1-dof}-\eqref{eq:P2-dof}
have been constructed in such a way that the orthogonality relationship
and decomposition of unity relations are fulfilled: \begin{subequations}
\begin{align}
\sum_{\beta}P_{ij}(I)_{\alpha\beta}P_{kl}(J)_{\beta\gamma} & =\delta_{jk}P_{il}(I)_{\alpha\gamma},\quad I,J=0,1,2\label{eq:orthogonality}\\
\sum_{i,J}P_{ii}(J)_{\alpha\beta} & =\delta_{\alpha\beta},\label{eq:decompositionUnity}\end{align}
 \end{subequations} where the Greek indices $\alpha,\beta,\gamma,\ldots$
represents here either tensorial, vector or no spatial indices.

\section{Analysing the Consistency of the Particle Spectrum for a General
Model\label{sec:Consistency-Analysis-of}}

To understand the convenience of orthonormal basis of operators for
the attainment of the propagators, let us work out a general spectral
analysis for models suitable with the symmetries defined by HLG models.
For such, let $\mathcal{L}_{2}$ be a quadratic Lagrangian for a HLG
model:

\begin{equation}
\mathcal{L}_{2}=\sum_{\alpha\beta}\varphi_{\alpha}\mathcal{O}_{\alpha\beta}\varphi_{\beta}+\sum_{\alpha}\varphi_{\alpha}\mathcal{S}_{\alpha},\label{eq:quad-lagr}\end{equation}
where the wave operator $\mathcal{O}_{\alpha\beta}$ is a local differential
operator, $\varphi_{\alpha}=\left\{ n,\ N_{a},\ h_{ab}\right\} $
carry the fundamental quantum fields of the model, and $\mathcal{S}_{\alpha}=\left\{ \mathcal{S},\mathcal{S}_{a},\mathcal{S}_{ab}\right\} $
are the matter sources of the lapse, shift, and spatial metric, respectively.
We can systematically analyze the spectrum and unitarity of this model
by means of a decomposition of the wave operator $\mathcal{O}_{\alpha\beta}$
in terms of the HLG projection operators $P_{ij}^{\varphi\psi}\left(J\right)$
in the momentum space:

\begin{equation}
\mathcal{L}_{2}=\sum_{\alpha\beta,ij,J}\varphi_{\alpha}a\left(J\right)_{ij}P_{ij}^{\varphi\psi}\left(J\right)_{\alpha\beta}\psi_{\beta},\label{eq:quad-lagr-proj}\end{equation}
where $a\left(J\right)_{ij}$ are the coefficients of the expansion
of the wave operator. The diagonal operators $P_{ii}^{\varphi\varphi}\left(J\right)$
are projectors in the spin-$J$ sector of the field $\varphi$ and
the off-diagonal operators $P_{ij}^{\varphi\psi}\left(J\right)$ $(i\neq j)$
implement mappings inside the spin subspace among the fields $\varphi$
and $\psi$.

The properties \eqref{eq:orthogonality}-\eqref{eq:decompositionUnity}
of the HLG projection operators reduces the task of inverting the
wave operator into a straightforward algebraic exercise. Indeed, if
the coefficient matrices $a(J)_{ij}$ are invertible, then the propagator
saturated with the external sources $\mathcal{S}_{\alpha}$ is given
by

\begin{equation}
\Pi=i\sum_{\alpha\beta,ij,J}\mathcal{S}_{\alpha}^{\ast}a\left(J\right)_{ij}^{-1}P_{ij}\left(J\right)_{\alpha\beta}\mathcal{S}_{\beta}'.\label{eq:propagatorInvertible}\end{equation}

Nevertheless, the coefficient matrices $a\left(J\right)_{ij}$ may
become degenerate due to a gauge symmetry of the Lagrangian model.
This poses no serious problem to the attainment of the propagator
since, in a such case, there are sources constraints that consistently
appear in order to inhibit the propagation of non-physical modes.
Both, the gauge transformations of the fields and the source constraints
may be obtained from the degeneracy structures of the coefficient
matrices,

\begin{subequations} \begin{align}
 & \delta\phi_{\alpha}={\displaystyle \sum_{\beta,J,i}V_{i}^{\left(R,n\right)}\left(J\right)P_{ij}\left(J\right)_{\alpha\beta}f_{\beta}\left(J\right),\ \ \text{for any \ensuremath{j}}}\label{eq:gauge-Symmetry}\\
 & \sum_{\beta,j}V_{j}^{\left(L,n\right)}\left(J\right)P_{ij}\left(J\right)_{\alpha\beta}\mathcal{S}_{\beta}'\left(J\right)=0,\ \ \text{for every }i\text{ and }J\label{eq:sourceConstraint}\end{align}
\end{subequations} with $f_{\beta}\left(J\right)$ being arbitrary
functions and $V^{\left(R,n\right)}$ ($V^{\left(L,n\right)}$) being
the right (left) null eigenvectors of $a\left(J\right)_{ij}$.

With it in mind, one should be convinced that the correct procedure
for the attainment of the propagator is to invert any largest nondegenerate
submatrix of $a\left(J\right)_{ij}$. In practice, it suffices to
delete the degenerate rows and columns of $a\left(J\right)_{ij}$
(which we denote $A\left(J\right)_{ij}$) according to the number
of gauge symmetries, to invert $A\left(J\right)_{ij}$, and saturate
it with physical sources, in order to obtain the propagator:

\begin{equation}
\Pi=i\sum_{\alpha\beta,ij,J}\mathcal{S}_{\alpha}^{*}A\left(J\right)_{ij}^{-1}P_{ij}\left(J\right)_{\alpha\beta}\mathcal{S}_{\beta}'.\label{eq:propagatorDegeneracies}\end{equation}

The propagators provides key ingredients for the understanding the
scattering processes of the theory under investigation. In particular,
we may analyze more carefully some conditions to guarantee the tree-level
unitarity and causality of the Lagrangian model. In general, the poles
of the propagators relation for HLG model is\begin{equation}
\omega^{2}-Q\left(k\right)=0,\end{equation}
where $Q$ may be a polynomial or rational function of the absolute
value of the spatial momentum $k$. So, for each propagating pole
we will require that $Q\left(k\right)\geq0$, for every momentum $k$,
to ensure it is not a tachyon mode. However, at low energies the higher
momentum terms becomes less relevant and the dispersion relation takes
a simple form \begin{equation}
\omega^{2}-c^{2}k^{2}=0.\end{equation}
So, for each propagating pole we will require that $c^{2}\geq0$ to
ensure it is not a tachyon mode.

On the other hand, the condition imposed by quantum field theory for
the absence of propagating ghosts reads

\begin{equation}
\Im m\text{Res}(\Pi|_{pole})>0.\label{eq:ghost conditions}\end{equation}
Fortunately, with the projection operators formalism we can take advantage
from the general decomposition of the degree-of-freedom HLG projection
operators,\begin{equation}
P_{ij}\left(J\right)_{\alpha\beta}=\left(-1\right)^{p}\psi_{\alpha}^{\left(i\right)}\psi_{\beta}^{\left(j\right)},\end{equation}
where $p$ is the parity related to the spin operator, to rewrite
the propagator \eqref{eq:propagatorInvertible} as\begin{eqnarray}
\Pi & = & i\left(-1\right)^{P}\sum_{J,ij,f}\mathcal{S}_{i}^{*}A\left(J,Q\right)_{ij}^{-1}\mathcal{S}_{j}'\left(\omega^{2}-Q\left(k\right)\right)^{-1},\end{eqnarray}
where $\mathcal{S}_{j}=\psi_{\alpha}^{\left(j\right)}\mathcal{S}^{\alpha}$
and $A\left(J,Q\right)_{ij}^{-1}$ are the inverse sub-matrices with
the pole extracted. Therefore, the positiveness condition \eqref{eq:ghost conditions},
for arbitrary sources, is ensured by the positiveness of the eigenvalues
of the $A\left(J,Q\right)_{ij}^{-1}$ matrix. Nevertheless, it can
be shown, for each case, that these matrices has only one non-vanishing
eigenvalue at the pole, which is equal to the trace of $A\left(J,Q\right)_{ij}^{-1}$.
Therefore, the condition for absence of ghosts for each spin is reduced
to:

\begin{equation}
(-1)^{P}\mbox{tr}A\left(J,Q\right)_{ij}^{-1}|_{pole}>0.\label{eq:ghost condition in terms of the trace}\end{equation}
This technical result converts the task of probing the unitarity,
that is in general a time-consuming work, into a straightforward algebraic
exercise of analyzing the matrices of coefficients.

\section{Description of the Lagrangian Model\label{sec:Description-of-the}}

Having discussed a general method to obtain the propagator and to
analyze the spectral consistency, we may focus our efforts in specific
HLG models of our interest. Since one can impose additional symmetries
and restrictions to the theory it must be made clear which version
of the theory one is referring to. In Ho\v{r}ava's original proposal,
probably concerned in maintaining its simplicity, he mainly discussed
the version with the \emph{detailed balance} condition. This condition
imposes that the possible spatial metric-dependent terms are of the
form $\mathcal{S}_{V}=\int dtd^{3}x\mathcal{N}\sqrt{g}\mathcal{E}^{ab}\mathcal{G}_{abcd}\mathcal{E}^{cd}$,
with $\mathcal{E}^{ab}$ itself following from a variational principle
$\sqrt{g}\mathcal{E}^{ab}=\frac{\delta W\left[g_{\cdot\cdot}\right]}{\delta g_{ab}}$.
Therefore, it implies in a huge reduction in the arbitrary parameters
of the model, however it has been shown it becomes very difficult
to reconcile this condition with phenomenological considerations \cite{Li:2009bg,Charmousis:2009tc,Blas:2009yd,Koyama:2009hc}.

Other requirement that significantly alters the structure of HLG models
is the \emph{projectability} condition. It enforces the lapse field
to be just a function of time $\mathcal{N}=\mathcal{N}\left(t\right)$.
Evidently, this avoids Lagrangian terms which are spatial derivatives
of $\mathcal{N}$, and thus simplifying the theory. A motivation for
imposing this assumption, comes from the possibility of setting $\mathcal{N}=1$
by a gauge fixing. In the following, we shall compare how this condition
affects the spectral properties of the model, but \emph{without} the
introduction of auxiliary fields.

Although interesting considerations can be made with both conditions,
we shall be mostly concerned with the version where the \emph{detailed
balance} and \emph{projectable }conditions are neglected. Allowing
numerous possible Lagrangian terms may turn the attainment of the
propagator a painful task, but with the HLG projection operators this
can be done with a systematic and simplified technique. This said,
let us enumerate the possible Lagrangian terms compatible with the
foliation-preserving diffeomorphisms (Eq. \eqref{eq:fDiffs}). 

First, we start with $K_{ab}K^{ab}$ and $K^{2}=\left(g^{ab}K_{ab}\right)^{2}$,
which are written in terms of the extrinsic curvature\begin{equation}
K_{ab}=\frac{1}{2\mathcal{N}}\left(\dot{g}_{ab}-\nabla_{a}N_{b}-\nabla_{b}N_{a}\right).\end{equation}
$K_{ab}$ is covariant under spatial diffeomorphisms and transforms
as a scalar under time reparametrization.  Here and throughout the
manuscript, the dot refers to time derivative and $\nabla_{a}$ is
the covariant derivative associated with the three-dimensional metric
$g_{ab}$. $K_{ab}K^{ab}$ and $K^{2}=\left(g^{ab}K_{ab}\right)^{2}$
are referred \emph{kinetic terms}, since they contain (at most two)
time derivatives.

The \emph{potential terms} do not contain time derivatives. The ones
built with the spatial metric $g_{ab}$ have been enumerated in Ref.
\cite{Sotiriou:2009bx}. Since in three-dimensions, the Riemann tensor
is equivalent to the Ricci tensor the independent possible terms up
to dimension $\left[\kappa\right]^{6}$ are $R^{3}$, $RR_{\ b}^{a}R_{\ a}^{b}$,
$R_{\ b}^{a}R_{\ c}^{b}R_{\ a}^{c}$, $R\nabla^{2}R$, $\nabla^{4}R$,
$\nabla_{a}R_{bc}\nabla^{a}R^{bc}$, $\varepsilon^{abc}R_{ad}D_{b}R_{c}^{\ d}$,
$R^{2}$, $R^{ab}R_{ab}$, $\nabla^{2}R$, $\varepsilon^{abc}\left(\Gamma_{ae}^{d}\partial_{b}\Gamma_{cd}^{e}+\frac{2}{3}\Gamma_{ad}^{f}\Gamma_{be}^{d}\Gamma_{cf}^{e}\right)$,
$R$, and 1.

However, the terms above do not exhaust all the possibilities. In
fact, an important novelty brought by Blas, Pujolás, and Sibiryakov
in Refs. \cite{Blas:2009qj,Blas:2010hb} was realizing that terms
involving the object $\mathcal{A}_{a}=\mathcal{N}^{-1}\partial_{a}\mathcal{N}$
should be included since it is invariant under foliation-preserving
diffeomorphism. The terms mixing $\mathcal{A}_{a}$ with the extrinsic
curvature $K_{ab}$ with dimension $\left[\kappa\right]^{6}$ or less
are $K^{ab}\mathcal{A}_{a}\mathcal{A}_{b}$, $K^{ab}\nabla_{a}\mathcal{A}_{b}$,
and $K\nabla_{a}\mathcal{A}^{a}$. They have the property to be odd
under T and CPT transformations. And the list of potential terms built
with $\mathcal{A}_{a}$ reads $\mathcal{A}_{a}\mathcal{A}^{a}$, $\mathcal{A}_{a}\nabla^{2}\mathcal{A}^{a}$,
$\left(\mathcal{A}_{a}\mathcal{A}^{a}\right)^{2}$, $\mathcal{A}_{a}\mathcal{A}_{b}R^{ab}$,$\mathcal{A}_{a}\nabla^{4}\mathcal{A}^{a}$,
$\left(\mathcal{A}_{a}\mathcal{A}^{a}\right)^{3}$, $\mathcal{A}_{a}\mathcal{A}^{a}\mathcal{A}_{b}\mathcal{A}_{c}R^{bc}$...
Notably, with this improvement it has been argued that HLG does not
run into problems with strong coupling and could be a phenomenologically
viable model \cite{Blas:2009qj,Blas:2010hb}.

After this brief consideration on the development of HLGs (for a deeper
discussion we address the reader to Refs. \cite{Horava:2009uw,Charmousis:2009tc,Sotiriou:2009bx,Blas:2009yd,Blas:2009qj,Weinfurtner:2010hz,Blas:2010hb,Padilla:2010ge,Sotiriou:2010wn}),
we set up a general HLG model only with the terms that affect the
propagator, \begin{equation}
S_{HL}=\int dtd^{3}x\sqrt{g}N\left(\alpha K_{ab}K^{ab}+\beta K^{2}+\mathcal{L}_{g}+\mathcal{L}_{\mathcal{A}}\right),\label{eq:actionHorava}\end{equation}
where the potential terms related only to the spatial metric are

\begin{subequations}\begin{equation}
\mathcal{L}_{g}=a_{1}R+b_{1}R_{ab}R^{ab}+c_{1}R^{2}+a_{2}\Delta R+b_{2}R_{ab}\Delta R^{ab}+c_{2}R\Delta R+a_{3}\Delta^{2}R+\mu\mathcal{L}_{CS}+\lambda\mathcal{L}_{RC},\end{equation}
where the Chern-Simons like term reads\begin{equation}
\mathcal{L}_{CS}=\varepsilon^{abc}\left(\Gamma_{ae}^{d}\partial_{b}\Gamma_{cd}^{e}+\frac{2}{3}\Gamma_{ad}^{f}\Gamma_{be}^{d}\Gamma_{cf}^{e}\right)\label{eq:chern-simons}\end{equation}
and the Ricci-Cotton term is cast as\begin{equation}
\mathcal{L}_{RC}=\varepsilon^{abc}R_{ad}D_{b}R_{c}^{\ d}.\label{eq:ricci-cotton}\end{equation}
The Lagrangian with the terms related with $\mathcal{A}_{a}$ reads,\begin{eqnarray}
\mathcal{L}_{\mathcal{A}} & = & rK^{ab}\nabla_{a}\mathcal{A}_{b}+sK\nabla_{a}\mathcal{A}^{a}+u_{1}R\nabla_{a}\mathcal{A}^{a}+u_{2}\Delta R\nabla_{a}\mathcal{A}^{a}\\
 & + & \eta_{1}\mathcal{A}_{a}\mathcal{A}^{a}+\eta_{2}\mathcal{A}_{a}\Delta\mathcal{A}^{a}+\eta_{3}\mathcal{A}_{a}\Delta^{2}\mathcal{A}^{a}.\nonumber \end{eqnarray}

\end{subequations}

We should emphasize that, at this point, we shall adopt the posture
of not avoiding the discussion of the terms that violate P, T or CPT
symmetry. Up until now, we have no fundamental reason for leaving
them out and it is our interest to understand how they affect the
propagating modes dispersion relations.

\section{Attainment of The Propagators and Spectral Analysis for the General
Ho\v{r}ava-Lifshitz Gravity model\label{sec:Attainment-of-The}}

Having established the model of our interest, we may work in the wave
operator defined by Eq. \eqref{eq:actionHorava}. After some algebraic
calculations (including partial integrations), it is possible to bring
the quadratic Lagrangian for the action \eqref{eq:actionHorava} into
the form of the Eq. \eqref{eq:quad-lagr}. With this procedure we
determine the corresponding wave operator. In the sequel, we perform
the wave operator decomposition in terms of the HLG projection operators,
as indicated by Eq. \eqref{eq:quad-lagr-proj}, aided by the relations
developed in the Appendix \ref{sec:Degree-of-Freedom-Projection-Operators}.

Some remarks are in order. In the attainment of the spin-1 matrix
of coefficients one realizes that there is no need for decomposing
the propagating modes its degree-of-freedom, since it appears no parity
violating contribution in this sector. For simplicity, we write them
in terms of the parity-preserving operators \eqref{eq:P1-spin}. On
the other hand, the Chern-Simons and Ricci-Cotton parity breaking
terms (Eqs. \eqref{eq:chern-simons} and \eqref{eq:ricci-cotton},
respectively) imply the violation of parity in the spin-2 sector.
It forces us to use the degree-of-freedom operators \eqref{eq:P2-dof}
in this sector. With this in mind, we write out the coefficients matrices
of the wave operator expansion,

\begin{subequations}

\begin{equation}
a\left(2\right)=\left(\begin{array}{cc}
\frac{\alpha}{2}\omega^{2}-\frac{1}{2}\left(a-bk^{2}\right)k^{2} & -ik^{2}\left(\mu+\frac{1}{2}\lambda k^{2}\right)\sqrt{k^{2}}\\
ik^{2}\left(\mu+\frac{1}{2}\lambda k^{2}\right)\sqrt{k^{2}} & \frac{\alpha}{2}\omega^{2}-\frac{1}{2}\left(a-bk^{2}\right)k^{2}\end{array}\right),\label{eq:coef-spin2}\end{equation}

\begin{equation}
a\left(1\right)=\left(\begin{array}{cc}
\frac{\alpha}{2}\omega^{2} & -\alpha\omega\frac{\sqrt{k^{2}}}{\sqrt{2}}\\
-\alpha\omega\frac{\sqrt{k^{2}}}{\sqrt{2}} & \alpha k^{2}\end{array}\right),\label{eq:coef-spin1}\end{equation}

\begin{equation}
a\left(0\right)=\left(\begin{array}{cccc}
\left(\frac{\alpha}{2}+\beta\right)\omega^{2}+\frac{1}{2}Ak^{2} & \frac{\beta}{\sqrt{2}}\omega^{2} & -\beta\omega\sqrt{2k^{2}} & \sqrt{2}\left(a-uk^{2}\right)k^{2}+\frac{1}{\sqrt{2}}si\omega k^{2}\\
\frac{\beta}{\sqrt{2}}\omega^{2} & \frac{1}{2}\left(\alpha+\beta\right)\omega^{2} & -\left(\alpha+\beta\right)\omega\sqrt{k^{2}} & \frac{1}{2}\left(r+s\right)i\omega k^{2}\\
-\beta\omega\sqrt{2k^{2}} & -\left(\alpha+\beta\right)\omega\sqrt{k^{2}} & 2\left(\alpha+\beta\right)k^{2} & -\left(r+s\right)ik^{2}\sqrt{k^{2}}\\
\sqrt{2}\left(a-uk^{2}\right)k^{2}-\frac{1}{\sqrt{2}}si\omega k^{2} & -\frac{1}{2}\left(r+s\right)i\omega k^{2} & \left(r+s\right)ik^{2}\sqrt{k^{2}} & 2\eta k^{2}\end{array}\right),\label{eq:coef-spin0}\end{equation}
\end{subequations}where we have redefined some coefficients to avoid
cluttering the notation: \begin{subequations}\begin{eqnarray}
a & = & a_{1}-a_{2}k^{2}+a_{3}k^{4},\\
b & = & b_{1}-b_{2}k^{2},\\
c & = & c_{1}-c_{2}k^{2},\\
u & = & u_{1}-u_{2}k^{2},\\
\eta & = & \eta_{1}-\eta_{2}k^{2}+\eta_{3}k^{4},\\
A & = & a+\left(3b+8c\right)k^{2}.\end{eqnarray}
\end{subequations}

As mentioned in Sec. \ref{sec:Decomposing-the-Graviton}, the degeneracy
structure of the matrices \eqref{eq:coef-spin2}-\eqref{eq:coef-spin0}
brings us information about the gauge symmetries and the source constraints
of the model. In particular, the spin-1 matrix is degenerate and the
$2^{nd}$ and $3^{rd}$ rows of the spin-0 matrix are proportional.
Adding together the result of equation \eqref{eq:sourceConstraint}
for these two matrices, one obtains\begin{equation}
\frac{\omega}{2}\mathcal{S}_{b}+k_{a}\mathcal{S}_{ab}=0.\label{eq:sourceConstraintHorava}\end{equation}
For a consistency test, one may verify that for the Einstein-Hilbert
gravity (EHG) ($\alpha=-\beta=a_{1}=\frac{1}{16\pi G}$ and all other
coefficients vanishing) there appears another symmetry in the spin-0
matrix: the $3^{rd}$ and $4^{th}$ columns and rows will become proportionate.
It will give rise to another constraint among the sources\begin{equation}
\omega\mathcal{S}+k_{a}\mathcal{S}_{a}=0\quad\mbox{(only for E-H)}.\label{eq:sourceConstraintEH}\end{equation}
These results are compatible with the relativistic source constraint
$k_{\mu}\mathcal{S}^{\mu\nu}=0$ ($\mu,\nu=0,1,2,3$ ) if one takes
into account that: (i) the contributions of $h^{0a}$ and $h^{a0}$
are encoded in $N^{a}$; it generates a factor of 2 in the matrix
decomposition, which implies in a $\frac{1}{2}$-factor in the source
$\mathcal{S}_{a}$ and (ii) the weak field expansion $\mathcal{N}=1+g$,
along with the definition in the ADM decomposition of $^{4}g_{00}$
implies that $h^{00}$ is identified with $2n$; this also brings
a $\frac{1}{2}$-factor associated with the source $\mathcal{S}$.

With these matrices coefficients, we may follow the analysis of the
spectral consistency as proposed in section \ref{sec:Decomposing-the-Graviton}.
For the sake of clarity, we separated the discussion for each spin
mode.

\subsection{Spin-2 Sector}

Let us analyze is the spin-2 mode of the graviton. The matrix $a\left(2\right)$
from equation \eqref{eq:coef-spin2} is non-degenerate. To obtain
the propagator \eqref{eq:propagatorInvertible} its inverse may be
readily obtained\begin{equation}
a\left(2\right)^{-1}=\frac{1}{D_{\left(2\right)}}\left(\begin{array}{cc}
\frac{\alpha}{2}\omega^{2}-\frac{1}{2}\left(a-bk^{2}\right)k^{2} & -i\left(\mu+\frac{1}{2}\lambda k^{2}\right)k^{2}\sqrt{k^{2}}\\
i\left(\mu+\frac{1}{2}\lambda k^{2}\right)k^{2}\sqrt{k^{2}} & \frac{\alpha}{2}\omega^{2}-\frac{1}{2}\left(a-bk^{2}\right)k^{2}\end{array}\right),\label{eq:propagator-spin2}\end{equation}
where the denominator $D_{\left(2\right)}=\left(\frac{\alpha}{2}\omega^{2}-\frac{1}{2}\left(a-bk^{2}\right)k^{2}\right)^{2}-\left(\mu+\frac{1}{2}\lambda k^{2}\right)^{2}k^{6}$
has two poles in the energy. This is an interesting feature of the
parity breaking terms: they allow each spin component to propagate
independently, with different dispersion relations\begin{equation}
\omega_{\pm}^{2}=\frac{1}{\alpha}\left(a-bk^{2}\right)k^{2}\pm\frac{2}{\alpha}\left(\mu+\frac{1}{2}\lambda k^{2}\right)k^{2}\sqrt{k^{2}}.\end{equation}

Using the posture of requiring a positive speed for every propagating
mode, one must ensure that \begin{equation}
c_{\left(2\right)}^{2}=\frac{a_{1}}{\alpha}>0.\end{equation}
 The condition for the absence of ghosts must be made for each pole
$\omega_{\pm}^{2}=Q_{\pm}\left(k\right)$. In either cases, \eqref{eq:ghost condition in terms of the trace}
implies that $1/\alpha>0$, so the ghost and tachyon free conditions
are given by\begin{equation}
\mbox{Spin-}\mathbf{2}:\ a_{1}>0;\ \alpha>0.\label{eq:spectral-spin-2}\end{equation}

From this analysis we conclude that the Chern-Simons and Ricci-Cotton
terms, which are related with the parameters $\mu$ and $\lambda$,
do not interfere in the unitary relations as its related terms in
the three-dimensional topological gravity.

\subsection{Spin-1 Sector}

Since the spin-1 matrix is degenerate, we shall use the prescription
of the equation \eqref{eq:propagatorDegeneracies}. A largest non-degenerate
matrix can be obtained by deleting its $2^{nd}$ columns and rows.
Its inverse is given by

\begin{equation}
A\left(1\right)^{-1}=\frac{2}{\alpha\omega^{2}}.\label{eq:propagator-spin1}\end{equation}

This mode poses us with a strange pole $\omega^{2}=0$. However, \eqref{eq:propagator-spin1}
is related to the projection operator $P_{11}^{hh}\left(1\right)=\frac{1}{2}\left(\theta_{ac}\omega_{bd}+\theta_{bc}\omega_{ad}+\theta_{ad}\omega_{bc}+\theta_{bd}\omega_{ac}\right)$,
which implies that the spatial momentum $k_{a}$ is contracted with
the source $\mathcal{S}^{ab}$ in the expression \eqref{eq:propagatorDegeneracies}.
In this way, using the source constraint \eqref{eq:sourceConstraintHorava},
one verifies that the residue at the pole of the propagator saturated
with physical sources vanishes identically. We must therefore understand
this pole as a non-propagating mode.

\subsection{Spin-0 Sector}

We left for last the spin-0 sector, which brings the most interesting
discussions in the literature. It is a natural consequence of the
lost of symmetry that new propagating modes appear and this is reflected
in this sector. However, it is of a major importance to verify the
spectral consistency of this mode. In fact, the EHG augmented with
relativistic higher-derivative corrections has been shown to be renormalizable
\cite{Stelle:1979}, but exactly the 4D spin 0 mode yields a ghost
excitation which jeopardizes unitarity of S-matrix and its possibility
to become a consistent quantum model for gravitation.

The second and third columns of the matrix $a\left(0\right)$ (equation
\eqref{eq:coef-spin0}) are degenerate, due to a residual gauge symmetry.
The inverse of $A\left(0\right)$ is cast in the Appendix \ref{sec:Inverse-of-A0}
and stems a dispersion relation,

\begin{equation}
\omega^{2}+\frac{\left(\frac{1}{2}A\eta-\left(a-uk^{2}\right)^{2}\right)\left(\alpha+\beta\right)-\frac{1}{8}Ak^{2}\left(r+s\right)^{2}}{\frac{1}{2}\alpha\eta\left(\alpha+3\beta\right)-\frac{1}{8}\left(\alpha\left(\left(r+s\right)^{2}+2s^{2}\right)+2\beta r^{2}\right)k^{2}}k^{2}=0.\end{equation}
This great number of arbitrary parameters makes the spectral analysis
a complicated tasks and many intriguing situation occurs. In the special
case where $\eta=u=0$, and $r=-s\neq0$ we obtain the dispersion
relation $\omega^{2}+4\frac{a^{2}}{r^{2}}=0$, which is a constant
negative energy solution.

If the CPT violating terms are absent $r=s=0$, we have no great difficulty
to proceed. In this situation,

\begin{equation}
A\left(0\right)^{-1}=\frac{1}{D_{\left(0\right)}}\left(\begin{array}{ccc}
2\eta\left(\alpha+\beta\right) & -2\sqrt{2}\eta\beta & -\sqrt{2}a\left(\alpha+\beta\right)\\
-2\sqrt{2}\eta\beta & \frac{2\eta\left(\alpha+2\beta\right)\omega^{2}-\left(4a^{2}-2A\eta\right)k^{2}}{\omega^{2}} & 2a\beta\\
-a\sqrt{2}\left(\alpha+\beta\right) & 2a\beta & \frac{\left(\alpha+3\beta\right)\alpha\omega^{2}+\left(\alpha+\beta\right)Ak^{2}}{2k^{2}}\end{array}\right),\label{eq:propagator-spin0}\end{equation}
with $D_{\left(0\right)}=\eta\left(\alpha+3\beta\right)\alpha\omega^{2}-\left(2a^{2}-A\eta\right)\left(\alpha+\beta\right)k^{2}$
and the dispersion relation is given by\begin{align}
\omega^{2}-\frac{\left(2a^{2}-A\eta\right)}{\eta}\frac{\left(\alpha+\beta\right)}{\left(\alpha+3\beta\right)\alpha}k^{2} & =0,\label{eq:dispersion-nonproj}\end{align}
which in general is a rational function of the spatial momentum $k^{2}$.
It recovers the dispersion relation obtained in the reference \cite{Blas:2009yd}. 

For small values of momentum the positivity of the propagation speed
for this mode reads,\begin{equation}
c_{\left(0\right)}^{2}=\frac{a_{1}\left(2a_{1}-\eta_{1}\right)}{\eta_{1}}\frac{\left(\alpha+\beta\right)}{\left(\alpha+3\beta\right)\alpha}>0.\label{eq:tachyon-cond-0}\end{equation}

To obtain the ghost-free condition, one ought to verify the residue
of the matrix at the pole. Replacing the dispersion relation \eqref{eq:dispersion-nonproj}
and after a inspection (in a CAS, for example) one verifies that
$A\left(0\right)^{-1}|_{pole}$ has only one non-vanishing eigenvalue,
which is equal to its trace, so the expression \eqref{eq:ghost condition in terms of the trace}
is valid,

\begin{eqnarray}
\mbox{tr} & A\left(0\right)^{-1}|_{pole}= & \frac{\left(a^{2}+2\eta^{2}\right)\left(\alpha+\beta\right)^{2}+4\beta^{2}\eta^{2}}{\alpha\eta^{2}\left(\alpha+\beta\right)\left(\alpha+3\beta\right)}.\end{eqnarray}

In this way, the condition for absence of ghosts can be stated as\begin{equation}
\alpha\left(\alpha+\beta\right)\left(\alpha+3\beta\right)>0.\label{eq:ghost-cond-0}\end{equation}
Combining the results \eqref{eq:tachyon-cond-0} and \eqref{eq:ghost-cond-0}
with the spin-2 condition \eqref{eq:spectral-spin-2} one must impose
that\begin{equation}
\mbox{Spin }\mathbf{0}:\ \left(\alpha+\beta\right)\left(\alpha+3\beta\right)>0;\ \eta_{1}\left(2a_{1}-\eta_{1}\right)>0,\end{equation}
for the absence of ghost and tachyon in this sector.

For comparison, let us come back and analyze how the \emph{projectability
condition} affects the matrix structures and thus the propagating
modes. Imposing the restriction that the lapse field does not depend
on the space variables $\mathcal{N}=\mathcal{N}\left(t\right)$ implies
that the terms in the quadratic Lagrangian containing the perturbation
of the lapse $n$ are total derivatives. Thus, the fourth column and
row of the spin 0 coefficient matrix vanishes identically, 

\begin{equation}
a\left(0\right)_{p}=\left(\begin{array}{cccc}
\left(\frac{\alpha}{2}+\beta\right)\omega^{2}+\frac{1}{2}Ak^{2} & \frac{\beta}{\sqrt{2}}\omega^{2} & -\beta\omega\sqrt{2k^{2}} & 0\\
\frac{\beta}{\sqrt{2}}\omega^{2} & \frac{1}{2}\left(\alpha+\beta\right)\omega^{2} & -\left(\alpha+\beta\right)\omega\sqrt{k^{2}} & 0\\
-\beta\omega\sqrt{2k^{2}} & -\left(\alpha+\beta\right)\omega\sqrt{k^{2}} & 2\left(\alpha+\beta\right)k^{2} & 0\\
0 & 0 & 0 & 0\end{array}\right).\end{equation}

With this simplification, one readily realizes that $a\left(0\right)_{p}$
is doubly degenerate: the column 4 is null and the columns 2 and 3
are proportional. A largest non-degenerate matrix of $a\left(0\right)_{p}$
is obtained by deleting the $3^{rd}$ and $4^{th}$ columns and rows,\begin{equation}
A\left(0\right)_{p}=\left(\begin{array}{cc}
\left(\frac{\alpha}{2}+\beta\right)\omega^{2}+\frac{1}{2}Ak^{2} & \frac{\beta}{\sqrt{2}}\omega^{2}\\
\frac{\beta}{\sqrt{2}}\omega^{2} & \frac{1}{2}\left(\alpha+\beta\right)\omega^{2}\end{array}\right).\end{equation}
Its inverse is given by\begin{equation}
A\left(0\right)_{p}^{-1}=\frac{4}{D_{\left(0\right)p}\omega^{2}}\left(\begin{array}{cc}
\frac{1}{2}\left(\alpha+\beta\right)\omega^{2} & -\frac{\beta}{\sqrt{2}}\omega^{2}\\
-\frac{\beta}{\sqrt{2}}\omega^{2} & \left(\frac{\alpha}{2}+\beta\right)\omega^{2}+\frac{1}{2}Ak^{2}\end{array}\right),\end{equation}
with\begin{equation}
D_{\left(0\right)p}=\left(\alpha+3\beta\right)\alpha\omega^{2}+\left(\alpha+\beta\right)Ak^{2}.\end{equation}
providing the dispersion relation for this propagating mode. To ensure
that this mode is not a tachyonic excitation we must impose\begin{equation}
c_{\left(0\right)p}^{2}=-\frac{\left(\alpha+\beta\right)a_{1}}{\left(\alpha+3\beta\right)\alpha}>0.\label{eq:tachyon-spin0-proj}\end{equation}

It can be explicitly verified that $\left.\mbox{Res}\left\{ A\left(0\right)^{-1}\right\} \right|_{pole}$
is degenerate, in such a way that its only non-vanishing eigenvalue
matches its trace. So one is enforced to impose that

\begin{align}
\mbox{tr}\left(\left.\mbox{Res}\left\{ A\left(0\right)^{-1}\right\} \right|_{pole}\right) & =\frac{2\left(\alpha+\beta\right)^{2}+4\beta^{2}}{\alpha\left(\alpha+\beta\right)\left(\alpha+3\beta\right)}>0.\label{eq:ghost-spin0-proj}\end{align}
The only way to reconcile the conditions \eqref{eq:tachyon-spin0-proj}
and \eqref{eq:ghost-spin0-proj} would be to impose $a_{1}<0$; but,
this contradicts the unitarity condition for the spin 2 sector. Therefore,
we may infer that the HLG with the \emph{projectable }condition is
not compatible with perturbative unitarity at the tree-level approximation.

An alternative to circumvent this issue is to enhance the model with
an Abelian gauge symmetry \cite{Horava:2010zj,Horava:2011gd,Borzou:2011mz,daSilva:2010bm},
in order to inhibit the propagation of the spin-0 mode . This is possible
with the introduction of a gauge field $A$ and an auxiliary scalar
field $\nu$, known as the Newtonian prepotential. This approach yields
a model with the same number of degrees-of-freedom as the EHG.

\section{Concluding Assessment\label{sec:Concluding-remarks}}

In this manuscript, we have built a set of orthonormal projection
operators suitable for non-relativistic theories of gravity, including
the ones with parity violating terms. The advantage of this construction
for the attainment of the propagator and the spectral analysis, which
is in general a time-consuming task, is that it can be done in a systematic
and simplified way, including the handling of the gauge symmetries
of the model. With this methodology, we were able to determine conditions
over the coefficients of general Ho\v{r}ava-Lifshitz gravity (HLG)
in order to restrict the propagation of tachyons and ghosts.

With the general result of the propagators (Eqs. \eqref{eq:propagator-spin2},
\eqref{eq:propagator-spin1}, and \eqref{eq:propagator-spin0}), we
may contemplate some results in a low-energy regime of the theory,
where the terms depending on higher spatial momentum become less relevant.
In such situation, the general action of the \emph{nonprojectable}
HLG\emph{ }(Eq. \eqref{eq:actionHorava}) may be simplified to

\begin{equation}
S_{HLG}=\int dtd^{3}x\sqrt{g}\mathcal{N}\left(\alpha K_{ab}K^{ab}+\beta K^{2}+aR+\eta\mathcal{A}_{a}\mathcal{A}^{a}\right).\end{equation}
and the inverse of the coefficient matrices of the propagator for
this simplified model are cast as

\begin{subequations}\begin{equation}
A\left(0\right)_{HLG}^{-1}=\frac{1}{D_{\left(0\right)}}\left(\begin{array}{ccc}
2\eta\left(\alpha+\beta\right) & -2\sqrt{2}\eta\beta & -\sqrt{2}a\left(\alpha+\beta\right)\\
-2\sqrt{2}\eta\beta & \frac{2\eta\left(\alpha+2\beta\right)\omega^{2}-2a\left(2a-\eta\right)k^{2}}{\omega^{2}} & 2a\beta\\
-\sqrt{2}a\left(\alpha+\beta\right) & 2a\beta & \frac{\left(\alpha+3\beta\right)\alpha\omega^{2}+\left(\alpha+\beta\right)ak^{2}}{2k^{2}}\end{array}\right),\label{eq:prop-low-ener-hor-spin0}\end{equation}

\begin{equation}
A\left(1\right)_{HLG}^{-1}=\frac{2}{\alpha\omega^{2}},\quad a\left(2\right)_{HLG}^{-1}=\frac{2}{\alpha\omega^{2}-ak^{2}},\label{eq:prop-low-ener-hor-spin1-2}\end{equation}
\end{subequations} with $D_{\left(0\right)}=\eta\left(\alpha+3\beta\right)\alpha\omega^{2}-a\left(2a-\eta\right)\left(\alpha+\beta\right)k^{2}$.

It is instructive to compare the propagators of the HLG (Eqs. \eqref{eq:prop-low-ener-hor-spin0}
and \eqref{eq:prop-low-ener-hor-spin1-2}) with the Einstein-Hilbert
gravity (EHG) propagators in ADM coordinates with HLG projection operators.
They may be written as

\begin{subequations}\begin{equation}
A\left(0\right)_{EHG}^{-1}=\frac{1}{\alpha k^{2}}\left(\begin{array}{cc}
0\  & \frac{1}{\sqrt{2}}\\
\frac{1}{\sqrt{2}}\  & \frac{1}{4k^{2}}\left(\omega^{2}-k^{2}\right)\end{array}\right),\label{eq::inverse-spin0-einstein}\end{equation}
\begin{equation}
A\left(1\right)_{EHG}^{-1}=\frac{2}{\alpha\omega^{2}},\quad a\left(2\right)_{EHG}^{-1}=\frac{2}{\alpha\left(\omega^{2}-k^{2}\right)}.\label{eq::inverse-spin1-2-einstein}\end{equation}
\end{subequations}

By a quick glance at these matrices of coefficients one may set up
comparisons between both models. Analyzing the pole structure of the
propagators, we may conclude that the spin-$2$ massless mode propagate
in both models, but with different speed of propagation. This corresponds
to the usual helicity-2 graviton mode of the relativistic model. Both
models coincide by not having dynamical poles in the spin-1 sector.
However, HLG has a propagating pole the spin-0 sector, which is absent
in the EHG.

This new degree of freedom may lead to odd effects. For example, the
helicity-0 mode of massless limit of Pauli-Fierz gravity (PFG) generates
a factor $\frac{3}{4}$ in the bending of the light by the Sun, if
one scales the coupling constant of both PFG and EHG such that the
Newtonian limit holds \cite{Iwasaki:1971uz,vanDam:1970vg,Zakharov:1970cc,Vainshtein:1972sx,VanNieuwenhuizen:1973qf}
(see Ref. \cite{Hinterbichler:2011tt} for a recent review on the
subject). This effect can be explained by the different way that the
helicity-2 and helicity-0 modes couple to scalar and vectorial matter
sources. One should be aware that this apparent paradox, known as
vDVZ discontinuity, appears only in the perturbative approach of the
linearized theory.

Now, we are ready to compare the HLG and EHG in a situation of a point
mass static source coupled to relativistic matter sources. The static
point-like source can be described by $\mathcal{S}=M\delta\left(\omega\right)$,
$\mathcal{S}_{a}=\mathcal{S}_{ab}=0$. The propagator amplitude $\Pi$
for this source coupled to arbitrary sources $\left\{ \mathcal{S}',\mathcal{S}_{a}',\mathcal{S}_{ab}'\right\} $
is \begin{subequations}\begin{eqnarray}
-i\Pi_{HLG} & = & M\delta\left(\omega\right)\frac{1}{\left(2a-\eta\right)k^{2}}\left\{ \left(\theta_{ab}-\frac{2\beta}{\left(\alpha+\beta\right)}\omega_{ab}\right)\mathcal{S}_{ab}'-\frac{1}{2}\mathcal{S}'\right\} ,\\
-i\Pi_{EHG} & = & M\delta\left(\omega\right)\frac{1}{2\alpha k^{2}}\left\{ \theta_{ab}\mathcal{S}_{ab}'-\frac{1}{2}\mathcal{S}'\right\} ,\end{eqnarray}
\end{subequations}respectively for HLG and EHG. In the particular
case of the Newtonian test with a point-like source of mass $M'$
separated by the distance of $R$, the energy of interaction is given
by \begin{equation}
E_{HLG}=-\frac{1}{2\left(2a-\eta\right)}\frac{MM'}{R},\quad E_{EHG}=-\frac{1}{4\alpha}\frac{MM'}{R},\label{eq:energyInteracScalar}\end{equation}
respectively for HLG and EHG.

On the other hand, the relativistic electromagnetic matter should
be dealt more carefully than just a scalar matter, for which $\mathcal{S}_{\ \mu}^{\mu}=0$.
In terms of the ADM-components the traceleness of the electromagnetic
tensor reads, \begin{equation}
\frac{1}{2}\mathcal{S}_{EM}'+\delta_{ab}\mathcal{S}_{EM}^{'ab}=0.\label{eq:EMsourceADMcond}\end{equation}
Using the identity \eqref{eq:EMsourceADMcond} together with the source
conservation constraint \eqref{eq:sourceConstraintHorava} one can
show that the interaction energy of the point mass with the light
beam exceeds the value of Eq. \eqref{eq:energyInteracScalar} by a
factor of two for both theories (replacing mass with total energy).
This implies that these experiments with static sources cannot distinguish
the EHG from the \emph{nonprojectable} HLG at the tree-level order
of perturbation, since both are scaled by multiplicative constant.

In this vein, one may look foward for a dynamical experiment that
may unveil the effects of the extra scalar mode. An interesting possibility
is to set up the test of \emph{gravitational pulsar} experiment. It
consists of a spherically symmetric harmonically pulsating mass distribution.
In fact, according to the Birkhoff theorem, the EHG radiation from
any spherically symmetric source is zero \cite{Anderson:1967}. Looking
to the propagators coefficients of the EHG (Eqs. \eqref{eq::inverse-spin0-einstein}-\eqref{eq::inverse-spin1-2-einstein})
one may verify that it is indeed the case, since the propagating pole
of the EHG is associated with a transverse operator $P\left(2\right)$
that vanishes when contracted with a spherically symmetric source.
The propagator structure of HLG leads to a different situation, because
it has non-vanishing coefficients with propagating pole associated
with longitudinal operators. In this case only the spin-0 mode carries
radiating energy and, hence, the Birkhoff theorem is not valid in
the \emph{nonprojectable} HLG.

Our efforts here do not contemplate the incorporation of the cosmological
constant in connection with the spectral analysis of the modes present
in HLG. We understand that this is a fairly interesting issue, which
may motivate further investigation. Indeed, in the recent paper of
Ref. \cite{Ong:2011km}, the authors pursue a stimulating discussion
on the stability of HLG black holes in connection with the AdS/CFT
correspondence and possible applications of the holographic principle.
In view of that, we shall be going further to endeavor a study of
the consequences of the cosmological constant in the frame of the
approach we report here.

Also, the coupling of Lifshitz fermions to HLG could be an interesting
matter in connection with our present investigation, since torsion
degrees of freedom may be excited and their incorporation shall yield
new interesting technical aspects in our treatment. The extension
of our method to include torsion in the HLG frame is also motivated
by the need to study the possible production and the consequent decay
of massive gravitons at the TeV scale to finally compare with the
results worked out in previous Lorentz-preserving quantum gravity
models \cite{Aquino:2011,Ask:2009,Gao:2010,Goldhaber:2010,Hagiwara:2008,Karg:2010,Ya-Jin:2007}.

\textbf{ACKNOWLEDGMENTS:}

The authors express their gratitude to Prof. A. J. Accioly for helpful
comments and suggestions. J. L. L. Morais, E. Scatena, and R. Turcati
are also acknowledged for the supporting discussions and G. N. Bremm
for the careful reading of the manuscript. Thanks are also due to
CNPq-Brazil and CAPES-Brazil for the financial support.

\appendix

\section{Inverse of $A\left(0\right)$ Matrix\label{sec:Inverse-of-A0}}

The inverse of $A\left(0\right)$ matrix that is obtained by deleting
the $3^{rd}$ row and column of $a\left(0\right)$ (equation \eqref{eq:coef-spin0})
is given by:

\begin{equation}
A\left(0\right)^{-1}=\frac{1}{D_{0}}\left(\begin{array}{ccc}
A_{11}^{\left(0\right)} & A_{12}^{\left(0\right)} & A_{13}^{\left(0\right)}\\
A_{12}^{\left(0\right)*} & A_{22}^{\left(0\right)} & A_{23}^{\left(0\right)}\\
A_{13}^{\left(0\right)*} & A_{23}^{\left(0\right)*} & A_{33}^{\left(0\right)}\end{array}\right),\label{eq:InvSpin0}\end{equation}
 where

\begin{subequations}\begin{eqnarray}
 & D_{0} & =\left\{ \left[\frac{1}{2}\alpha\eta\left(\alpha+3\beta\right)-\frac{1}{8}k^{2}\left(\alpha\left(\left(r+s\right)^{2}+2s^{2}\right)+2\beta r^{2}\right)\right]\omega^{2}\right.\label{eq:den0}\\
 &  & \left.+\left[\left(\frac{1}{2}A\eta-\left(a-uk^{2}\right)^{2}\right)\left(\alpha+\beta\right)-\frac{1}{8}Ak^{2}\left(r+s\right)^{2}\right]k^{2}\right\} k^{2}\omega^{2}\nonumber \\
 & A_{11}^{\left(0\right)} & =\left(\left(\alpha+\beta\right)\eta-\frac{1}{4}\left(r+s\right)^{2}k^{2}\right)\omega^{2}k^{2}\\
 & A_{12}^{\left(0\right)} & =-\sqrt{2}\beta\eta\omega^{2}k^{2}-\frac{i}{\sqrt{2}}\left(r+s\right)\left(a-uk^{2}+\frac{1}{2}si\omega\right)\omega k^{4}\\
 & A_{13}^{\left(0\right)} & =-\left(\frac{1}{2\sqrt{2}}i\left(s\alpha-r\beta\right)\omega+\frac{1}{\sqrt{2}}\left(a-uk^{2}\right)\left(\alpha+\beta\right)\right)\omega^{2}k^{2}\\
 & A_{22}^{\left(0\right)} & =\left(\alpha+2\beta\right)\eta\omega^{2}k^{2}-\left(\frac{1}{2}s^{2}\omega^{2}+2\left(a-uk^{2}\right)^{2}-A\eta\right)k^{4}\\
 & A_{23}^{\left(0\right)} & =-\frac{1}{4}i\left(\left(r+s\right)\alpha+2r\beta\right)k^{2}\omega^{3}+\left(a-uk^{2}\right)\beta k^{2}\omega^{2}-\frac{1}{4}iAk^{4}\left(r+s\right)\omega\\
 & A_{33}^{\left(0\right)} & =\frac{1}{4}\left(\left(\alpha+3\beta\right)\alpha\omega^{2}+A\left(\alpha+\beta\right)k^{2}\right)\omega^{2}\end{eqnarray}
\end{subequations}

{\small }{\small \par}

\section{Ho\v{r}ava-Lifshitz Gravity Projection Operators and Tensorial Relations\label{sec:Degree-of-Freedom-Projection-Operators}}

In order to facilitate further use of the set of the degree-of-freedom
and the parity preserving HLG projection operators we write out some
useful identities satisfied by them. To avoid cluttering the notation,
some indices of operators may be omitted.

\subsection{\emph{Vector field operators:} $N^{a}-N^{a}$}

\paragraph*{Identities Among the Operators}

\begin{equation}
P^{NN}\left(1\right)_{ab}=P_{33}^{NN}\left(1^{++}\right)_{ab}+P_{44}^{NN}\left(1^{--}\right)_{ab}\end{equation}

\paragraph*{Tensorial Identities}

\begin{subequations}

\begin{eqnarray}
 & \delta_{ab} & =P_{33}^{NN}\left(0\right)_{ab}+P^{NN}\left(1\right)_{ab}\\
 & k_{a}k_{b} & =k^{2}P_{33}^{NN}\left(0\right)_{ab}\\
 & \varepsilon_{abc}k^{c} & =\sqrt{k^{2}}\left(P_{12}^{NN}\left(1^{++}\right)_{ab}-P_{21}^{NN}\left(1^{--}\right)_{ab}\right)\end{eqnarray}

\end{subequations}

\subsection{Rank-2 Symmetric Field Operators, $h-h$ }

\paragraph*{Identities Among the Operators}

\begin{subequations}\begin{eqnarray}
P^{hh}\left(1\right)_{ab;cd} & = & P_{11}^{hh}\left(1^{++}\right)+P_{22}^{hh}\left(1^{--}\right)\\
P^{hh}\left(2\right)_{ab;cd} & = & P_{11}^{hh}\left(2^{++}\right)+P_{22}^{hh}\left(2^{--}\right)\end{eqnarray}
\end{subequations}

\paragraph*{Tensorial Identities}

\begin{subequations}

\begin{eqnarray}
 &  & \delta_{ab,cd}=\frac{1}{2}\left(\delta_{ac}\delta_{bd}+\delta_{ad}\delta_{bc}\right)=P^{hh}\left(2\right)+P^{hh}\left(1\right)+P_{11}^{hh}\left(0^{s}\right)+P_{22}^{hh}\left(0^{\omega}\right)\\
 &  & \delta_{ab}\delta_{cd}=2P_{11}^{hh}\left(0^{s}\right)+\sqrt{2}P_{12}^{hh}\left(0^{s\omega}\right)+\sqrt{2}P_{21}^{hh}\left(0^{\omega s}\right)+P_{22}^{hh}\left(0^{\omega}\right)\\
 &  & k_{a}k_{b}\delta_{cd}+k_{c}k_{d}\delta_{ab}=\sqrt{2}k^{2}\left(P_{12}^{hh}\left(0^{s\omega}\right)+P_{21}^{hh}\left(0^{\omega s}\right)\right)+2k^{2}P_{22}^{hh}\left(0^{\omega}\right)\\
 &  & k_{a}k_{c}\delta_{bd}+k_{a}k_{d}\delta_{bc}+k_{b}k_{c}\delta_{ad}+k_{b}k_{d}\delta_{ac}=2k^{2}P^{hh}\left(1\right)+4k^{2}P_{22}^{hh}\left(0^{\omega}\right)\\
 &  & k_{a}k_{b}k_{c}k_{d}=k^{4}P_{22}^{hh}\left(0^{\omega}\right)\\
 &  & \left(\varepsilon_{eca}\delta_{bd}+\varepsilon_{ecb}\delta_{ad}+\varepsilon_{eda}\delta_{bc}+\varepsilon_{edb}\delta_{ac}\right)k^{e}=2\sqrt{k^{2}}\left(2P_{12}^{hh}\left(2^{-+}\right)-2P_{21}^{hh}\left(2^{+-}\right)-P_{12}^{hh}\left(1^{+-}\right)+P_{21}^{hh}\left(1^{-+}\right)\right)\\
 &  & \left(\varepsilon_{eca}k_{b}k_{d}+\varepsilon_{ecb}k_{a}k_{d}+\varepsilon_{eda}k_{b}k_{c}+\varepsilon_{edb}k_{a}k_{c}\right)k^{e}=2k^{2}\sqrt{k^{2}}\left(-P_{12}^{hh}\left(1^{+-}\right)+P_{21}^{hh}\left(1^{-+}\right)\right)\end{eqnarray}

\end{subequations}

\subsection{Graviton-Vector Operator, $h-N$ }

\paragraph*{Tensorial Identities}

\begin{subequations}\begin{eqnarray}
 &  & P^{hN}\left(1^{+}\right)_{ab;c}=P_{13}^{hN}\left(1^{++}\right)+P_{24}^{hN}\left(1^{--}\right)\\
 &  & \delta_{ab}k_{c}=\sqrt{2k^{2}}P_{13}^{hN}\left(0^{s}\right)+\sqrt{k^{2}}P_{23}^{hN}\left(0^{++}\right)\\
 &  & \frac{1}{2}\left(k_{a}\delta_{cb}+k_{b}\delta_{ca}\right)=\frac{\sqrt{k^{2}}}{\sqrt{2}}\left(P_{13}^{hN}\left(1^{++}\right)+P_{24}^{hN}\left(1^{--}\right)\right)+\sqrt{k^{2}}P_{23}^{hN}\left(0^{++}\right)\end{eqnarray}
\end{subequations}

\subsection{Graviton-Vector Operator, $N-h$}

\paragraph*{Tensorial Identities}

\begin{subequations}\begin{eqnarray}
 &  & P^{Nh}\left(1^{+}\right)_{a;bc}=P_{31}^{Nh}\left(1^{++}\right)+P_{42}^{Nh}\left(1^{--}\right)\\
 &  & k_{a}\delta_{bc}=\sqrt{2k^{2}}P_{31}^{Nh}\left(0^{s}\right)+\sqrt{k^{2}}P_{32}^{Nh}\left(0^{\omega}\right)\\
 &  & \frac{1}{2}\left(k_{b}\delta_{ac}+k_{c}\delta_{ab}\right)=\frac{\sqrt{k^{2}}}{\sqrt{2}}\left(P_{31}^{Nh}\left(1^{++}\right)+P_{42}^{Nh}\left(1^{--}\right)\right)+\sqrt{k^{2}}P_{32}^{Nh}\left(0^{\omega}\right)\end{eqnarray}
\end{subequations}

\end{document}